\documentclass[a4paper,conference]{IEEEtran}
\usepackage{graphicx}
\usepackage{amsmath}
\usepackage{algorithm2e}
\usepackage{multirow}
\usepackage{url}

\IEEEoverridecommandlockouts

\ifCLASSINFOpdf
  
\else
  
\fi

\def\pol{{\mbox{\footnotesize pol}}}

\begin{document}

\title{Quantitative Assessment of TV White Space in India\thanks{This work
has been supported by the Ford Foundation.}}

\author{\IEEEauthorblockN{Gaurang Naik, Sudesh Singhal, Animesh Kumar, and
Abhay Karandikar}
\IEEEauthorblockA{Department of Electrical Engineering\\
Indian Institute of Technology Bombay\\
Mumbai - 400076\\
Email: \{gaurangnaik,sudesh,animesh,karandi\}@ee.iitb.ac.in}
}

\maketitle

\begin{abstract}
Licensed but unutilized television (TV) band spectrum is called as TV white
space in the literature. Ultra high frequency (UHF) TV band spectrum has
very good wireless radio propagation characteristics.  The amount of TV
white space in the UHF TV band in India is of interest.
Comprehensive quantitative assessment and estimates for the TV white space
in the $470$-$590$MHz band for four zones of India (all except north) are
presented in this work. This is the first effort in India to estimate TV
white spaces in a comprehensive manner.  The average available TV white
space per unit area in these four zones is calculated using two methods:
(i) the primary (licensed) user and secondary (unlicensed) user point of
view; and, (ii) the regulations of Federal Communications Commission in the
United States.  By both methods, the average available TV white space in
the UHF TV band is shown to be more than $100$MHz!  A TV transmitter
frequency-reassignment algorithm is also described. Based on spatial-reuse
ideas, a TV channel allocation scheme is presented which results in
insignicant interference to the TV receivers while using the least number
of TV channels for transmission across the four zones.  Based on this
reassignment, it is found that four TV band channels (or $32$MHz) are
sufficient to provide the existing UHF TV band coverage in India.
\end{abstract}

\IEEEpeerreviewmaketitle

\section{Introduction}

With rising demand for bandwidth, several researchers around the world
have measured and studied the occupancy of spectrum in different
countries.  These measurements suggest that except for the spectrum
allocated to services like cellular technologies, and the industrial,
scientific and medical (ISM) bands, most of the allocated spectrum is
heavily underutilized. The overall usage of the analyzed spectrum is
as low as 4.54\% in Singapore \cite{spectrum2}, 6.2\% in Auckland
\cite{spectrum1}, 17.4\% in Chicago \cite{spectrum4} and 22.57\% in
Barcelona \cite{spectrum3}.  Among all the unutilized portions of the
frequency spectrum, white spaces in the Ultra High Frequency (UHF)
Television (TV) bands have been of particular interest owing to the
superior propagation characteristics as compared to the higher
frequency bands. 

Loosely speaking, the unutilized (or underutilized) TV channels
collectively form the TV white spaces. The amount of available TV
white space varies with location and time.  TV white space estimation
has been done in countries like the United States (US), the United
Kingdom (UK), Europe, and
Japan~\cite{quant_3,quant_UK,quant_europe,quant_Japan}. In the Indian
context, single-day experiments at three locations in urban and
sub-urban Delhi have been performed~\cite{iitd}. The estimation of TV
white space in the UHF band, based on spectrum allocation and TV
transmitter parameters, is presented in this work.

The main contributions of this paper are the following:
\begin{enumerate}
 \item For the first time, the empirical quantification of the
available TV white space in the $470$-$590$MHz in India is presented.
The quantification utilizes existing methods in the literature, namely
pollution and protection viewpoints \cite{quant_3}, and the technical
specifications of the Federal Communications
Commission~\cite{FCC_2008}. It is found that UHF TV band spectrum is
heavily underutilized in India.
\item Motivated by underutilization of UHF TV band spectrum, a spatial
reuse based channel allocation algorithm has been proposed for the
existing Indian TV transmitters operating in the 470-590 MHz band.
The algorithm uses the least number of TV channels while ensuring no
(significant) interference between transmitters operating in the same
channel. It is observed that at least $70\%$ UHF TV band channels can
be freed by this approach.
\end{enumerate}

The importance of the above results must be understood in the context of
Indian National Frequency Allocation Plan (NFAP) 2011 where a policy intent
for the utilization of TV white spaces was made.  Therefore, it is
necessary to estimate the amount of TV white spaces in India.  Besides,
based on above results, the TV band in India is underutilized and this
situation is quite different than in the developed countries.  The optimal
mechanism(s) for the use of TV white spaces in India can be
\textit{different} and it should be studied by further research.

\textit{Organization:} The TV white space scenario and the related
work on quantitative analysis in a few other countries is briefly
described in Sec.~\ref{sec:tvws_review}.
Sec.~\ref{sec:india_tvplan} describes the current Indian usage
scenario of the UHF TV Bands.  Sec.~\ref{sec:methodology} presents the methodology and
assumptions used in calculating the white space availability in India.
Sec.~\ref{sec:results} presents the results of our work, and compares the TV white
space availability in India with that of other countries.  In
Sec.~\ref{sec:channel_allocation}, we
propose a frequency allocation scheme to the TV transmitters in India so as
to ensure minimum number of channel usage in the country.  Concluding
remarks and directions for future work are discussed in
Sec.~\ref{sec:conclusions}.

\section{TV White Space in other countries}
\label{sec:tvws_review}

Regulators FCC in the US and Ofcom in the UK have allowed for
secondary operations in the TV white spaces. Under this provision, a
secondary user can use the unutilized TV spectrum provided it does not
cause harmful interference to the TV band users and it relinquishes
the spectrum when a primary user (such as TV Transmitter) starts
operation. Since the actual availability of TV white spaces varies
both with location and time, operators of secondary services are
interested in the amount of available white space.  The available TV
white space depends on regulations such as the protection margin to
the primary user, maximum height above average terrain (HAAT),
transmission power of secondary user, and the separation distance.

As per FCC, a band can declared as unutilized if no primary signal is
detected above a threshold of $-114$dBm~\cite{FCC_2008}. Using the
parameters of terrestrial TV towers, TV white space availability in the US
has been done in the literature~\cite{quant_3}. The average number of
channels available per user has been calculated using the pollution and
protection viewpoints.\footnote{The pollution viewpoint looks at TV white
space calculations from the secondary users' point of view, whereas the
protection viewpoint is concerned with avoiding interference to the primary
users~\cite{quant_3}.} These viewpoints are explained in more detail in
Sec.~\ref{sec:methodology}.  Using the pollution viewpoint into account,
the average number of channels available per location increases with the
allowable pollution level. This average number of available channels is
maximum in the lower UHF band. In the protection viewpoint too, the average
number of available channels at a location is maximum in the lower UHF band
(channels 14-51 of the US) and this decreases as more and more constraints
are applied. In UK, Ofcom published a consultation providing details of
cognitive access to TV white spaces in 2009~\cite{Ofcom_2}.  The coverage
maps and database of digital TV (DTV) transmitters can be used to develop a
method for identification of the TV white space at any location within
UK~\cite{quant_UK}.  The TV white space availability in Japan has also been
studied in \cite{quant_Japan}.  The results of \cite{quant_Japan} indicate
that the amount of available TV white space in Japan is larger than that in
US and UK.  However, this availability decreases with an increase in the
separation distance.

To the best of our knowledge, a comprehensive study of TV white space
availability has not been done in India and is the focus of this work.

% The number of available TV channels for secondary usage vary considerably
% from one location to another and is extremely low in regions with high
% population density.
% Moreover, as the adjacent channel constraint is applied, the number of
% available TV channels for secondary usage decreases further.  Using -114
% dBm FCC ruling, very little white space can be extracted in UK.

\section{Current Indian TV Band Plan}
\label{sec:india_tvplan}

As per the NFAP 2011
\cite{NFAP}, the spectrum in the frequency band $470$-$890$MHz is earmarked
for Fixed, Mobile and Broadcasting Services.  The NFAP has allowed the
digital broadcasting services to operate in the $585$-$698$MHz band.  India
is a part of the ITU Region~3, and the $698$-$806$MHz band has been
earmarked for International Mobile Telecommunications-Advanced (IMT-A)
applications (see footnote IND~38 of \cite{NFAP}).  Hence, the digital TV
broadcasting will operate in the frequency band from $585$MHz to $698$MHz.
Currently the TV transmitters operate only in the $470$-$590$MHz band in
the UHF band.

In India, the sole terrestrial TV service provider is Doordarshan which
currently transmits in two channels in most parts across India.  Currently
Doordarshan has $1415$ TV transmitters operating in India, out of which $8$
transmitters transmit in the VHF Band-I ($54$-$68$MHz comprising of two
channels of $7$MHz each), $1034$ transmitters operate in the VHF Band-III
($174$-$230$MHz comprising of eight channels of $8$MHz each), and the
remaining $373$ transmitters transmit in the UHF Band-IV ($470$-$590$MHz
comprising of fifteen channels of $8$MHz each).  In India, a small number
of transmitters operate in the UHF bands. As a result, apart from
$8$-$16$MHz band depending on the location, \textit{the UHF band is quite
sparsely utilized in India}! This observation will be made more precise in
the next section. 

\section{Methodology}
\label{sec:methodology}

The quantification of TV white space in India will be addressed in
this section. A computational tool has been developed that calculates
the protection region and separation distance for each tower, and also
the pollution region around the tower where a secondary device should
not operate. Currently, there are no TV white space regulations in
India. The regulations of FCC (US) are borrowed for the estimation of
TV white space in India. \textit{Microphones are ignored in our
computation due to lack of available information}.  The input to the
developed computational tool include the following parameters for all
the TV transmitters:
\begin{enumerate}
\item position of the tower (latitude and longitude),
\item transmission power of the TV transmitter,
\item frequency of operation,
\item height of the antenna,
\item and, terrain information of area surrounding the tower.
\end{enumerate}
The above parameters of all the TV towers operating in the UHF band-IV
have been obtained from the terrestrial TV broadcaster
Doordarshan.\footnote{While the TV tower data is available online
publicly in US and other countries, it is not so in India.  We could
obtain the data with considerable efforts from Doordarshan.}  Out of
373 transmitters operating in the 470-590 MHz UHF band (Channel No.
21-35), Doordarshan has provided data of 254 towers operating in the
West, East, South and the North East zone.  The TV transmitter
information for the North zone is yet to be provided by Doordarshan.
% In \cite{quant_Japan} and \cite{Ofcom_2}, the propagation model used
% is the ITU-R P.1546 model. However, the Indian terrain conditions
% are significantly different from those of the western countries.

Comprehensive field strength measurements in India suggest that Hata model
is fairly accurate for propagation modeling~\cite{comp,hata}. The Hata
model will be used for path-loss calculations.  Using the TV transmitter
information and the propagation model, we quantify the available TV white
space in the UHF TV band by two methods. The first method utilizes the
protection and pollution viewpoints while the second one utilizes
technical specification made by the FCC.

\subsection{Method 1: the protection and pollution viewpoints}

The protection and pollution viewpoints used for calculating TV white
space have been introduced by Mishra and Sahai~\cite{quant_3}. Their
method is reviewed in this section and utilized in our work for
obtaining TV white space availability (see Sec.~\ref{sec:results}).

\subsubsection{Protection viewpoint}

\begin{figure}[h]
\centering
\includegraphics[width=2.5in]{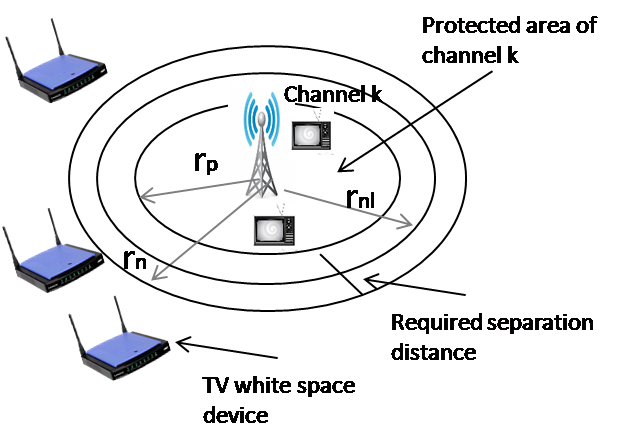}
\caption{\label{fig:picture31} Protection radius, separation distance
and the no-talk radius}
\end{figure}

In the protection viewpoint, when a secondary user operates, it must
not cause any interference to the primary receivers in its vicinity.
This is illustrated in Fig.~\ref{fig:picture31}
The protected area is defined using the following $\mbox{SINR}$
equations. Let $P_{t}$ be the transmit power of primary in dBm,
$PL(r)$ be the path-loss in dB at a radial distance $r$ from the
transmitter, $N_0$ be the thermal noise in dBm, and $\Delta$ be the
threshold $\mbox{SINR}$ in dBm. Then, the protection radius $r_p$ is
defined by the following $\mbox{SINR}$ equation,
\begin{align*}
P_{t} - PL(r_p) - N_{0} = \Delta.
\end{align*}
The regulator provides an additional margin ($\Psi$) to account for
fading. The modified equation for $r_p$ is, 
\begin{align*}
P_{t} - PL(r_p) - N_{0} = \Delta+\Psi.
\end{align*}
The no-talk radius $r_{n}$ is defined as the distance from the
transmitter up to which no secondary user can transmit.  The
difference $r_{n}-r_{p}$ is calculated such that if a secondary device
transmits at a distance of $r_{n}-r_{p}$ from the TV band receiver
located at $r_{p}$, the $\mbox{SINR}$ at the TV band receiver within a radius
$r_n$ does not fall below $\Delta$.\footnote{For simplicity, only one
secondary device transmitting around the primary receiver is
considered.} The separation distance $r_{n}-r_{p}$ is then calculated
such that
\begin{align*}
P_{s}-PL(r_{n}-r_{p}) = \Psi,
\end{align*}
where, $P_{s}$ is the secondary transmitter power in dBm.

In addition to the co-channel considerations, a TV receiver tuned to a
particular channel has a tolerance limit on the interference level in the
adjacent bands.  In the protection viewpoint, we consider that the
protection radius in the adjacent channel is the same as in co-channel.
However, the TV receiver can tolerate more adjacent channel interference
than co-channel interference.  Therefore, a margin of $27$dB more than
co-channel fading margin $\Psi$ (set by the FCC regulations
\cite{FCC_2008}) is provisioned for adjacent channel interference.

\subsubsection{Pollution viewpoint}

The pollution viewpoint takes into consideration the fact that even
though a region could be used by a secondary device, the interference
at the secondary receiver due to the primary transmitter might be
higher than the tolerable interference level of the secondary
receiver.  If $\gamma$ is the interference tolerable by the secondary
receiver, then $r_{\pol}$ is given by,
\begin{align*}
P_{t}-PL(r_{\pol}) = N_{0} + \gamma.
\end{align*}
Similar to the protection viewpoint, there are adjacent channel
conditions (leakage of primary transmitter's power in the adjacent
channel) in the pollution viewpoint as well.  It is assumed that the
secondary device can tolerate up to $45$dB of interference if it is
operating in the adjacent channels.  The TV white space available is
the \textit{intersection} of the white space determined from the
pollution and protection viewpoints.

The parameters used in our computations for calculating the available
TV white space are given in Table~\ref{I}.
\begin{table}[h]
\caption{Parameters used for calculation of TV white space using
pollution and protection viewpoints}
\label{I}
\centering
\begin{tabular}{|p{5.8cm}|p{2.1cm}|}
\hline
\multicolumn{2}{|c|}{\textbf{Pollution Viewpoint}}\\
\hline
\multirow{2}{*}{Maximum tolerable interference ($\gamma$) by
secondary} & $5$dB\\
& $15$dB (specified for 802.11g systems)\\
\hline
Maximum tolerable interference ($\gamma$) by secondary (adjacent
channel) & $45$dB\\
\hline
Noise in a $8$MHz band ($N_{0}$) & $-104.97$dBm \\
\hline
\multicolumn{2}{|c|}{\textbf{Protection Viewpoint}}\\
\hline
\multirow{2}{*}{Target fading margin ($\Psi$)} & $0.1$dB\\
& $1$dB (specified by FCC)\\
\hline
Additional fading margin in adjacent channel & $27$dB (specified by
FCC)\\
\hline
Required $\mbox{SINR}$ for primary receiver & $45$dB\\
\hline
Transmission power of secondary device & $36$dBm\\ 
\hline
HAAT of secondary device & $30$m\\
\hline
\end{tabular}
\end{table}
As an example, we consider the TV tower located at the Sinhagad Fort
in Pune.  Doordarshan informed us that the tower at Sinhagad Fort
operates in the $534$-$542$MHz band (channel~$29$) at a height of
$100$m and power of $10$kW ($70$dBm).  In the Hata model used for path
loss calculations, Pune has been considered as an urban city. Using
the pollution viewpoint, for a $15$dB tolerable interference in
channel $29$ ($534$-$542$MHz), the pollution radius for the tower is
calculated to be $37.70$km, and for a tolerable interference of $45$dB
in the adjacent channel, the pollution radius is $4.24$km.  What this
means for a secondary device is that the interference level is more
than the allowable limit ($15$dB above noise floor) in a region of
$37.70$km in channel~$29$ and $4.24$km in the adjacent channels around
the tower.

From the protection viewpoint, if a fading margin of $1$dB is
provided, the protection and no-talk radius in channel~$29$ are
$33.82$km and $33.83$km respectively.  If we consider an additional
fading margin of $27$dB in the adjacent band, the no talk radius in
the adjacent channel is $33.82$km.  This implies that if a secondary
device operates within a distance of $33.83$km in channel~$29$ and
$33.82$km in the adjacent channels, the primary user receiving on
channel~$29$ will experience interference.  The available white space
is the intersection of the white space using the two viewpoints.
Thus, in Pune, no secondary device can operate within a distance of
$37.70$km (limit set by pollution viewpoint) on channel~$29$ and
$33.82$ km in the adjacent channels (limit set by protection
viewpoint) around the tower at Sinhagad Fort.

\subsection{Method 2: TV white space calculation using FCC rules}
In the FCC's definition of TV white space, the protection radius is
same in the Grade B contour ($r_{b}$) \cite{quant_3, FCC_2008}.  In the
UHF band, $r_{b}$ is the distance from the TV tower where the field
strength of the primary signal falls to $41$dBu.  The required field
strength is converted from from dBu to dBm using the following
conversion formula \cite{oet},
\begin{align*}
P(\text{dBm}) &= E(\text{dBu}) - 130.8 + 20 \log_{10} \left(
\frac{1230}{f_H + f_L} \right),
\end{align*}
where, $P(\text{dBm})$ is transmit power in dBm, $E(\text{dBu})$  is
the field strength in dBu, $f_H$ is the upper frequency-limit of the
channel, and $f_L$ is the lower frequency-limit of the channel.  To
calculate the separation distance, i.e. distance beyond $r_{b}$ where
no secondary device can transmit, the distance $r_{n}-r_{b}$ such that
the signal from the secondary device at $r_{n}$ results in a signal
level of $E_{r_{b}} -23$dBu at the TV receiver located at $r_{b}$ is
calculated.  For the TV transmitter at Pune, the no-talk radius, i.e.
the distance from the tower beyond which a secondary device can use
the channel is computed to be $41.60$km.

\section{Results}
\label{sec:results}

The results obtained by TV white space calculation methods of
Sec.~\ref{sec:methodology} will be discussed in this section.

\subsection{White Space Availability using Pollution, Protection viewpoints \& the FCC Rule}

\begin{figure}[h]
\centering
\includegraphics[trim = 95mm 35mm 55mm 10mm, clip, width=3.6in]{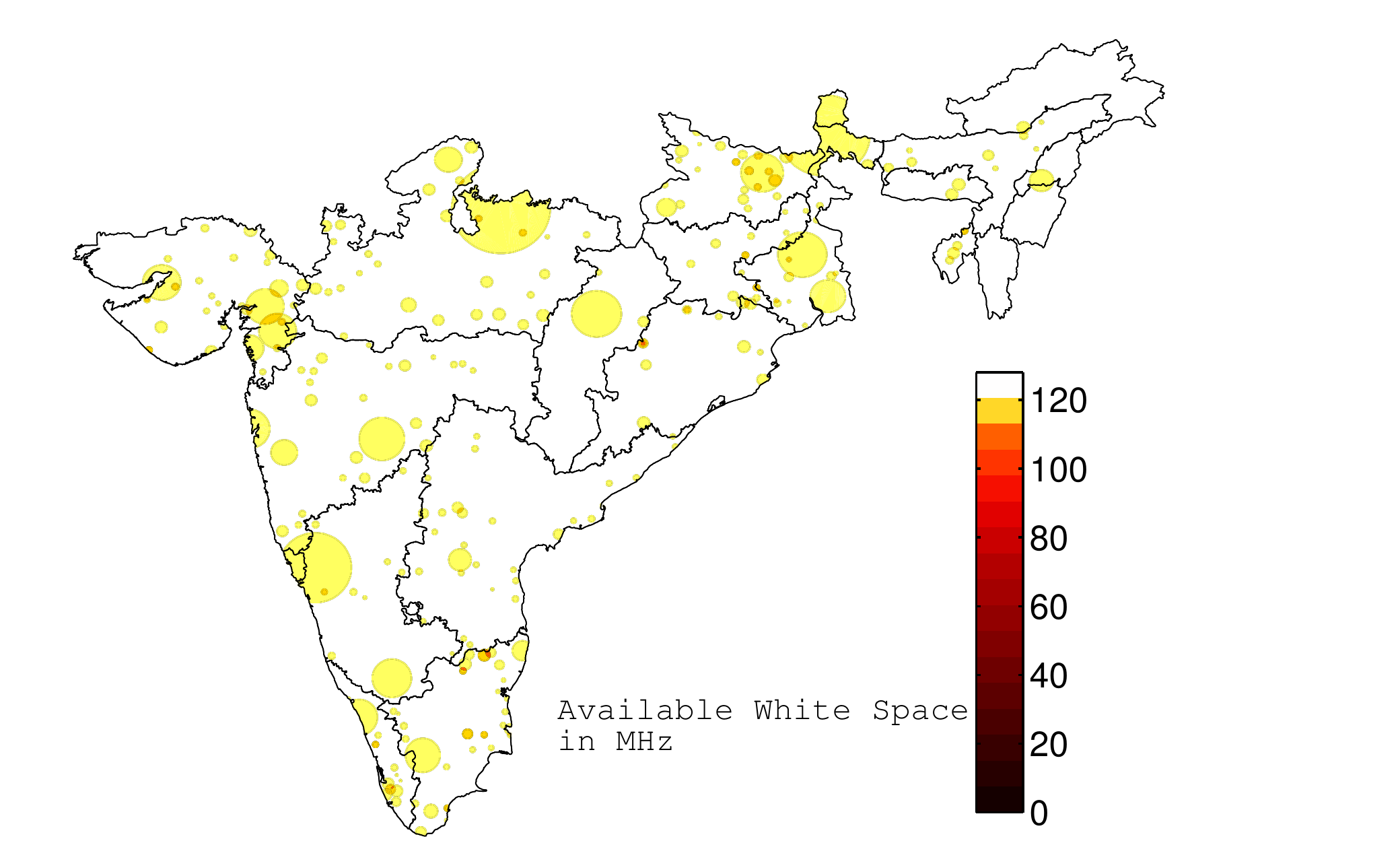}
\caption{TV White Space availability using Pollution viewpoint $\gamma =
15dB$}
\label{poll15db}
\end{figure}

\begin{figure}[h]
\centering
\includegraphics[trim = 95mm 35mm 55mm 10mm, clip, width=3.6in]{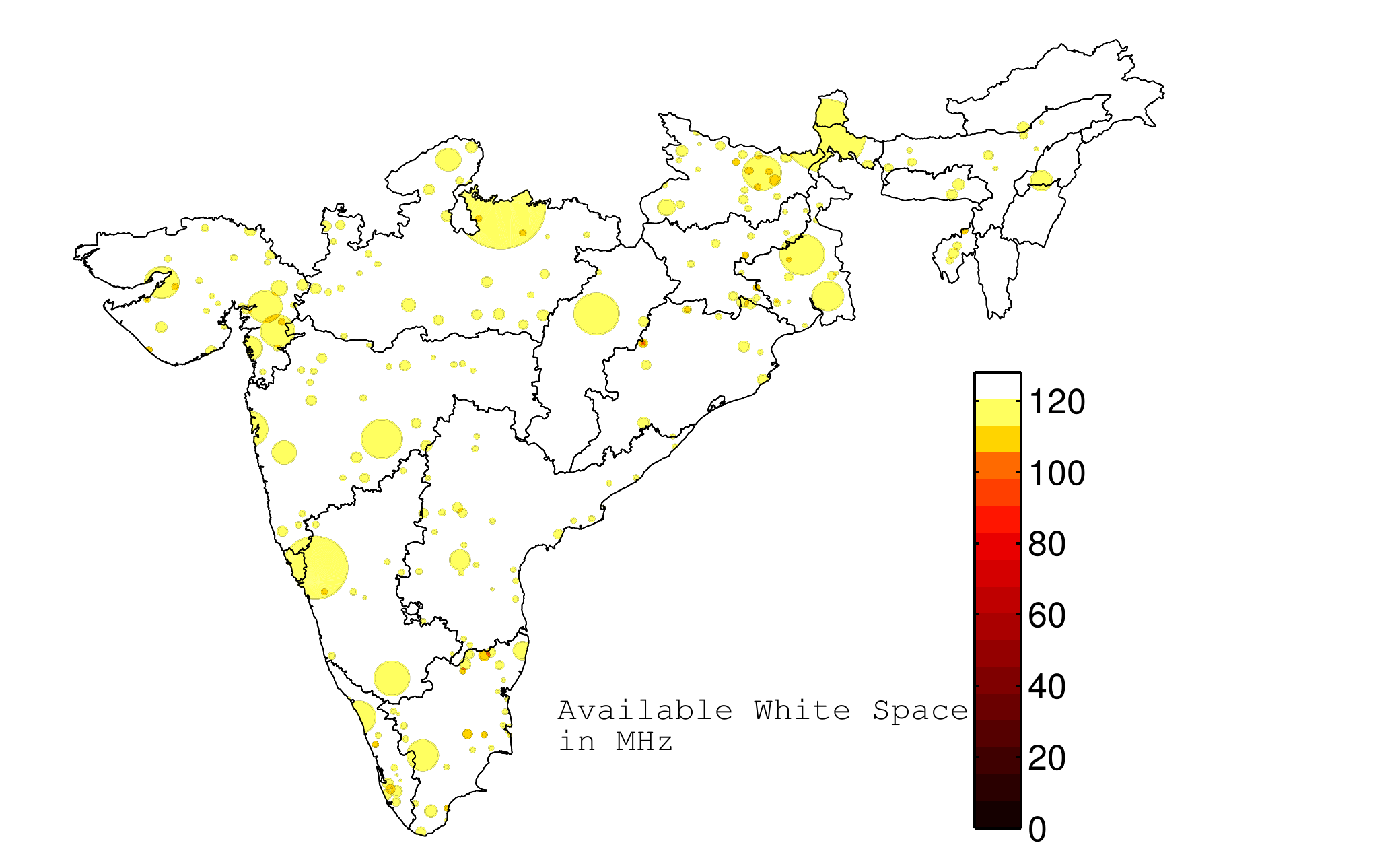}
\caption{TV White Space availability using Protection viewpoint $\Delta =
1dB$}
\label{prot1db}
\end{figure}

\begin{figure}[h]
\centering
\includegraphics[trim = 95mm 35mm 55mm 10mm, clip, width=3.6in]{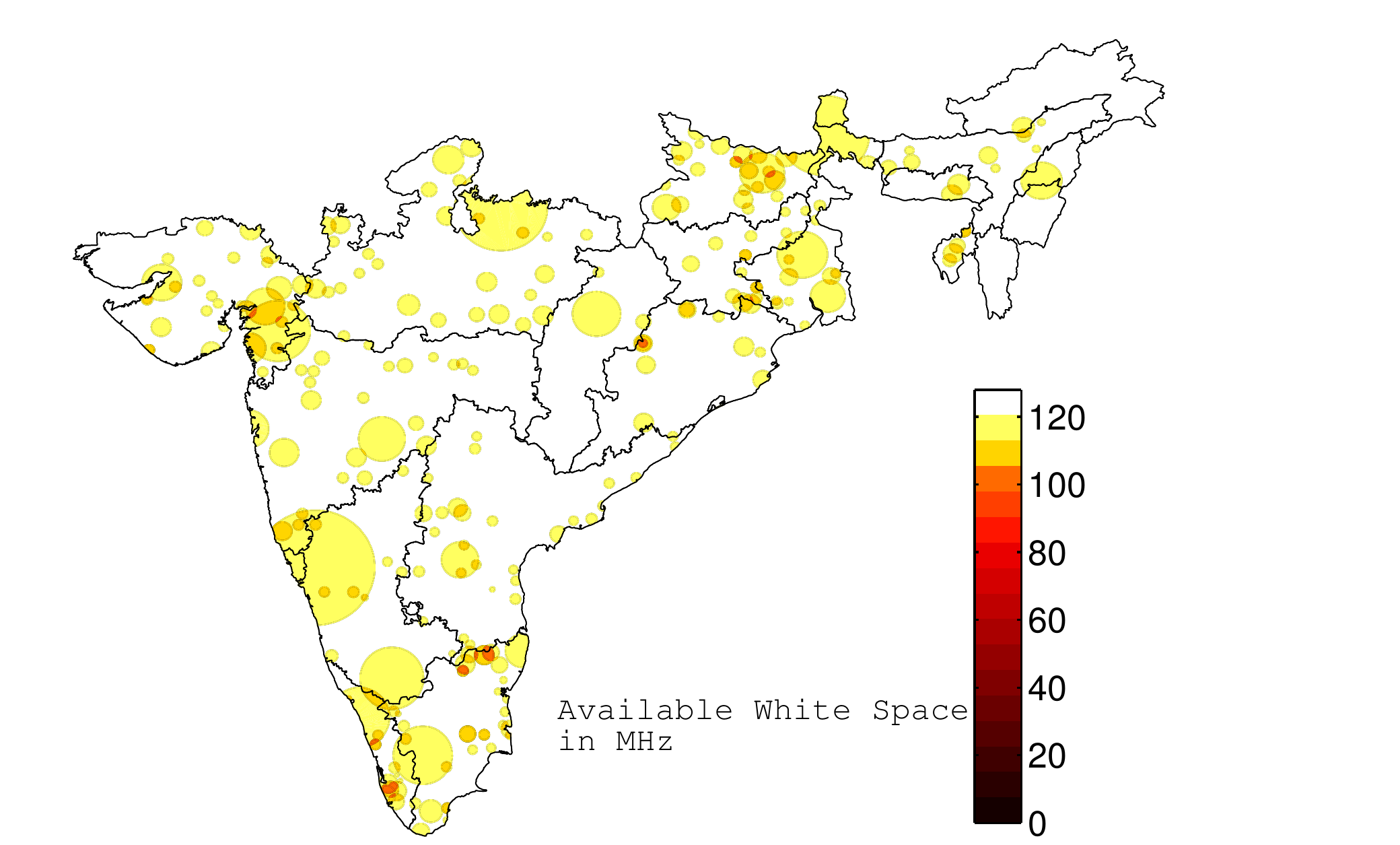}
\caption{TV White Space availability using FCC rule}
\label{fcc2}
\end{figure}

Using the methodology described in Sec.~III, the pollution and the
no-talk radius are calculated for every TV tower in the four zones.  Each
region is plotted as a circle around the TV tower. Here it has been assumed
that each tower has an omnidirectional antenna.  The TV white space
availability in the west, east, south and north-east zone using the
pollution viewpoint is shown in Fig.~\ref{poll15db} and using the
protection viewpoint in Fig.~\ref{prot1db}.  White space availability
using the FCC regulations is shown in Fig.~\ref{fcc2}.

\begin{figure}[h]
\centering
\includegraphics[trim = 25mm 0mm 25mm 0mm, clip, width=3.2in]{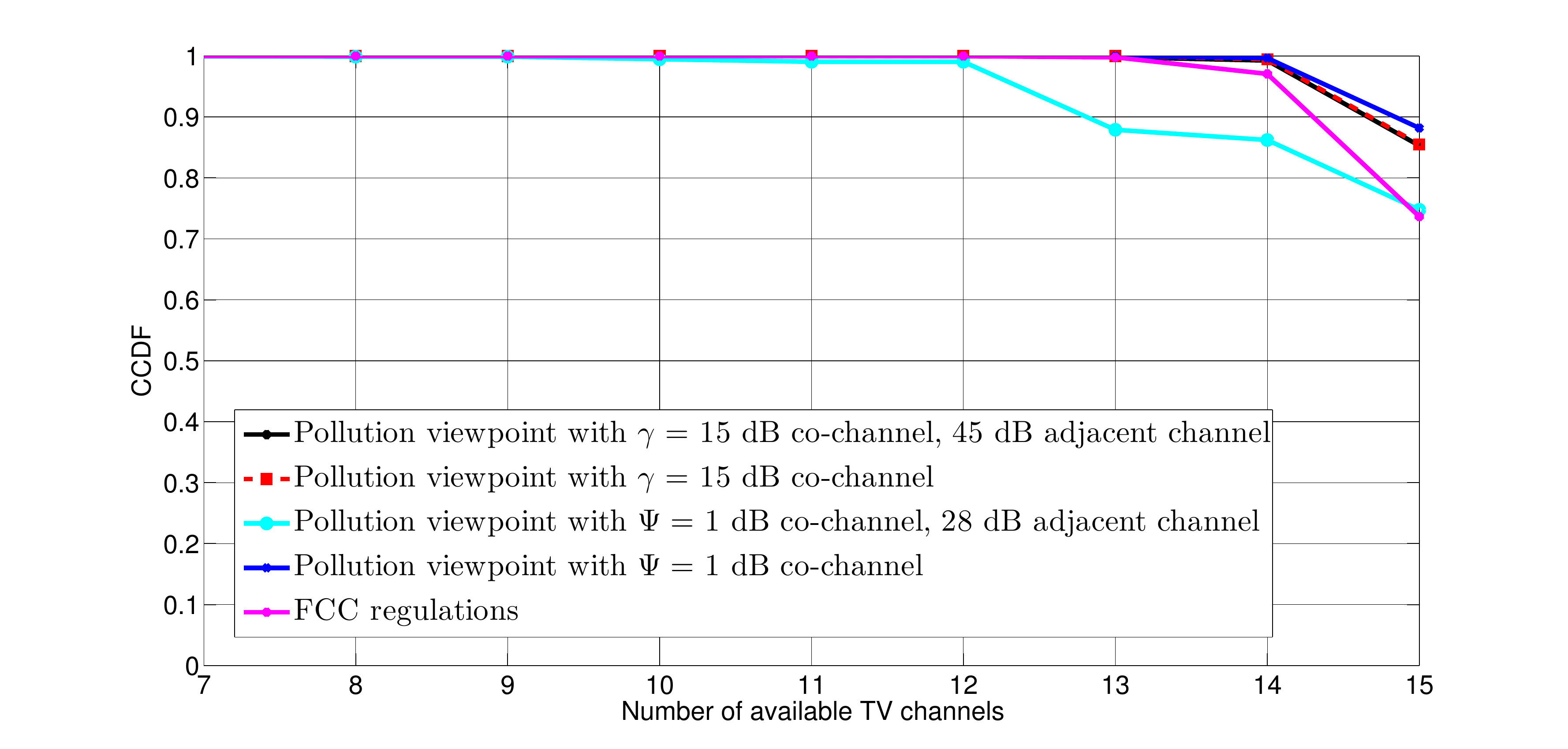}
\caption{Complementary cumulative distribution of the available number of
channels per area as white space}
\label{ccdf1}
\end{figure}

Fig.~\ref{poll15db} and Fig.~\ref{prot1db} illustrate that at most places
in India, not even a single channel in the UHF band is utilized!  To
quantify this result further, the complementary cumulative distribution
function of the number of channels available per unit area as TV white
space is plotted in Fig.~\ref{ccdf1}.
%
% \begin{figure}[h] \centering \includegraphics[trim = 25mm 0mm 25mm 0mm,
% clip, width=3.5in]{ccdf31.png} \caption{CCDF of available number of
% channels per area as white space} \label{ccdf1} \end{figure}
%
\begin{table*}[!htb]
\caption{Average number of channels available per unit area in each zone
(out of $15$ channels)}
\label{II}
\centering
\begin{tabular}{|c|l|p{1cm}|p{1cm}|p{1cm}|p{1cm}|p{1cm}|}
\hline
\textbf{Method} & \textbf{Parameters} & \textbf{West zone} & \textbf{East
zone} & \textbf{South zone} & \textbf{North east zone} & \textbf{India}\\
\hline
\multirow{6}{*}{Pollution Viewpoint} & Main channel $\gamma$ = 5 dB &
14.0047 & 14.2313 & 14.4745 & 14.8464 & 14.2130\\
& Main channel $\gamma$ = 5 dB, Adjacent channel $\gamma$ = 45 dB & 13.9957
& 14.2223 & 14.4693 & 14.8443 & 14.2054\\
& Main channel $\gamma$ = 10 dB & 14.6896 & 14.6295 & 14.7374 & 14.9213 &
14.6856\\
& Main channel $\gamma$ = 10 dB, Adjacent channel $\gamma$ = 45 dB &
14.6835 & 14.6205 & 14.7322 & 14.9192 & 14.6792\\
& Main channel $\gamma$ = 15 dB & 14.8545 & 14.8214 & 14.8683 & 14.9599 &
14.8496\\
& Main channel $\gamma$ = 15 dB, Adjacent channel $\gamma$ = 45 dB &
14.8485 & 14.8123 & 14.8630 & 14.9578 & 14.8432\\
\hline
\multirow{4}{*}{Protection Viewpoint} & Main channel $\Psi$ = 1 dB &
14.8830 & 14.8549 & 14.8917 & 14.9673 & 14.8782\\
& Main channel $\Psi$ = 1 dB, Adjacent channel $\Psi$ = 28 dB & 14.5372 &
14.4558 & 14.3666 & 14.6939 & 14.4616\\
& Main channel $\Psi$ = 0.1 dB & 14.8664 & 14.8351 & 14.8782 & 14.9630 &
14.8616\\
& Main channel $\Psi$ = 0.1 dB, Adjacent channel $\Psi$ = 27.1 dB & 14.4720
& 14.8429 & 14.5745 & 14.8661 & 14.4792\\
\hline
FCC regulations & Main channel $E_{r_{b}}$ = 41 dBu & 14.7762 & 14.6795 &
14.6510 & 14.8844 & 14.7050\\
\hline
\end{tabular}
\end{table*}
From the pollution viewpoint with $\gamma$ = $15$dB, 85.29\% of the area in
India has all the 15 channels available as white space, while in 100\% of
the area, 12 or more channels are available as white space.  Similar
results are obtained using the protection viewpoint with $\Psi$ = 1 dB
which shows that in 74.75\% of the area in India, all the 15 channels are
available for TV white space secondary operations, and in 99.47\% of the
area, 10 or more channels are available as white space.  With the FCC
regulations, which are considered to be conservative (see~\cite{quant_3}),
73.66\% of the area in India have all 15 channels available for TV white
space operations, while in 100\% of the area 12 or more channels are
available as white space.  Table~\ref{II} gives the average number of
channels available in the UHF TV bands in the four zones using different
methods described earlier.
\begin{table*}[!htb]
\caption{Number of available TV channels as a function of percentage area}
\label{III}
\centering
\begin{tabular}{|c|l|p{1.9cm}|p{1.9cm}|p{1.9cm}|}
\hline
 \textbf{Method} & \textbf{Parameters} & \textbf{10 channels free} &
\textbf{12 channels free} &  \textbf{15 channels free} \\
\hline
\multirow{6}{*}{Pollution Viewpoint} & Main channel $\gamma$ = 5 dB & 100\%
& 100\% &  36.69\%\\
& Main channel $\gamma$ = 5 dB, adjacent channel $\gamma$ = 45 dB & 100\% &
100\% &  36.43\% \\
& Main channel $\gamma$ = 10 dB & 100\% & 100\% & 71.61\%\\
& Main channel $\gamma$ = 10 dB, adjacent channel $\gamma$ = 45 dB & 100\%
& 100\% & 71.39\% \\
& Main channel $\gamma$ = 15 dB & 100\% & 100\% & 85.51 \\
& Main channel $\gamma$ = 15 dB, adjacent channel $\gamma$ = 45 dB & 100\%
& 100\%  & 85.29\% \\
\hline
\multirow{4}{*}{Protection viewpoint} & Main channel $\Psi$ = 1 dB & 100\%
& 100\% & 88.19\% \\
& Main channel $\Psi$ = 1 dB, adjacent channel $\Psi$ = 28 dB & 99.88\% &
99.04\% & 74.75\% \\
& Main channel $\Psi$ = 0.1 dB & 100\% & 100\% & 86.62\%\\
& Main channel $\Psi$ = 0.1 dB, Adjacent channel $\Psi$ = 27.1 dB & 99.99\%
& 99.57\% & 73.69\% \\
\hline
FCC Regulations & Main channel $E_{r_{b}}$ =41 dBu & 100\% & 100\% &
73.66\% \\
\hline
\end{tabular}
\end{table*}
Conclusions that can be drawn from Table~\ref{II} are as follows:
\begin{enumerate}
 \item Out of the 15 UHF TV channels (470-590 MHz), the
average number of TV channels available for secondary usage is above 14
(112MHz) in each of the four zones.
\item Available TV white space is the maxium in the North East,
where 18 transmitters operate in the UHF band.
 \item 
% In the pollution viewpoint, as the value of $\gamma$ increases i.e. the
% allowable pollution in the co-channel increases, the available white
% space increases.
If we use the adjacent channel constraint, the available
white space decreases. However, this decrease is less than 1\% in each
case.
%
% \item
% Using the protection viewpoint, as $\Psi$ increases, the available white
% space also increases.  The is because, as $\Psi$ increases, the secondary
% device can operate closer to the primary device operating at $r_{p}$.
% The value of $\Psi$ is decided by the regulator, and thus
% availability of white space using the protection viewpoint depends on the
% regulations.
%
\end{enumerate}

\subsection{Comparison of Indian TV white space Scenario with other countries}

Table \ref{III} concludes that in almost all cases at least 12 out of the
15 channels (80\%) are available as TV white space in 100\% of the areas in
India.  This is larger than Japan \cite{quant_Japan}, where out of 40
channels, on an average 16.67 channels (41.67\%) are available in 84.3\% of
the areas.  This white space is also larger than what is available in US
and the European countries.  The available TV white space by area in
Germany, UK, Switzerland, Denmark on an average are 19.2 (48\%), 23.1
(58\%), 25.3 (63\%) and 24.4 (61\%) channels out of the 40 channels
respectively \cite{quant_europe}.  Similarly, as compared to the US, the
available TV white space in India is much larger. It must be noted that in
TV white space studies across the world, the IMT-A band is also included.
%  Moreover, in all the studies  considered in the other countries, 40
%  channels have been considered for TV white space calculations. This
%  includes the Higher UHF band (channels~$52$-$69$). This band has been
%  earmarked for International Mobile Telecommunications (IMT) applications
%  in most of the regions (698-960 MHz in Region 2 and 790-960 MHz in
%  Regions 1 \& 3) as per ITU Radio Regulations Edition of 2012 \cite{ITU}.
%  As a result, the available TV white space is expected to further reduce.
%  On the other hand, in India, we have calculated the available TV white
%  space only in the 470-590 MHz band, and yet more than 100 MHz of
%  spectrum is available.

\section{Proposed Channel Allocation Scheme}
\label{sec:channel_allocation}

\begin{figure}[h]
\centering
\includegraphics[trim = 75mm 0mm 85mm 20mm, clip, width=3.3in]{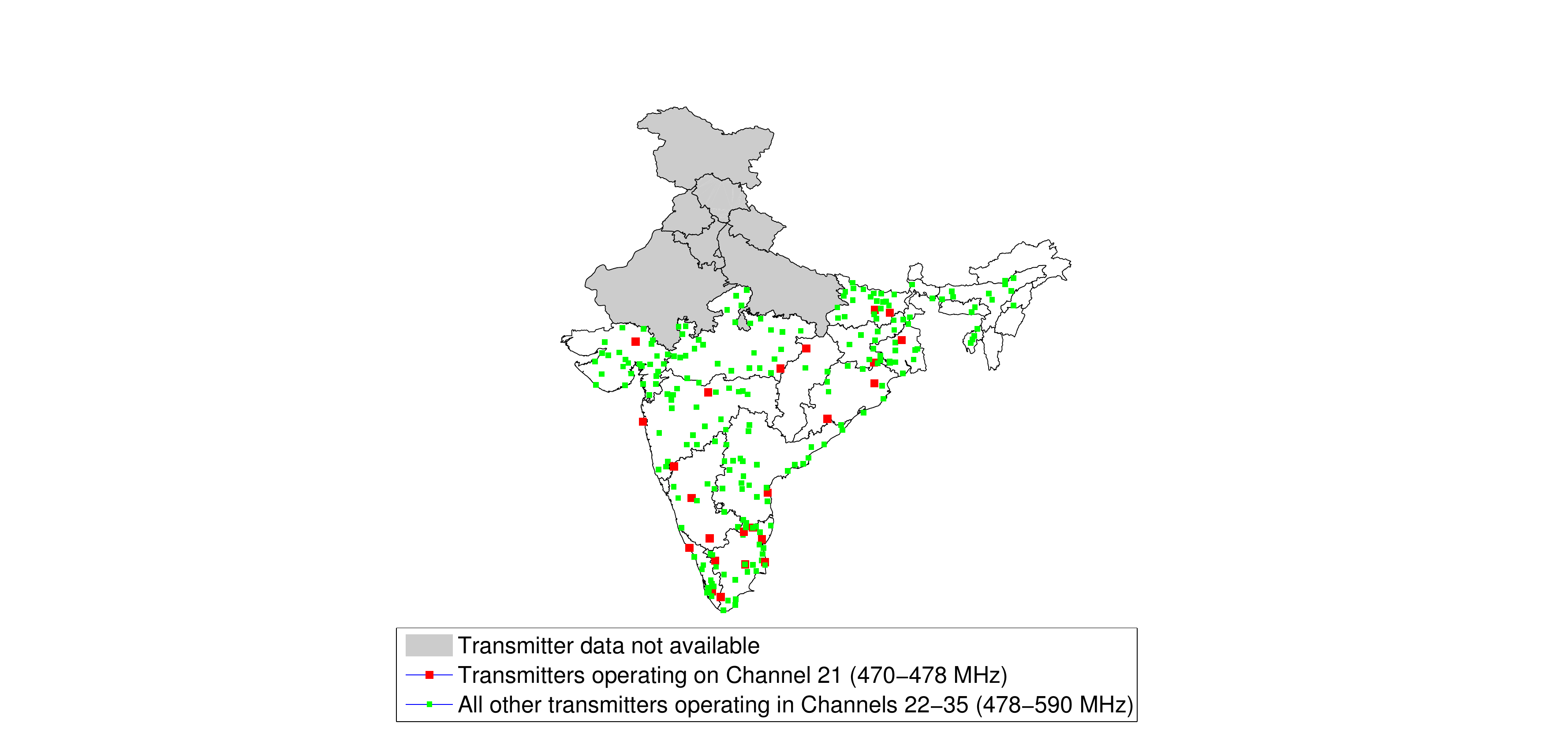}
\caption{TV Transmitters operating in UHF Band-IV ($470$-$590$MHz)}
\label{loc}
\end{figure}

There are a total of 254 Doordarshan TV transmitters in the four zones
illustrated in Fig.~\ref{loc} operating in the $470$-$590$MHz.  Currently,
in these zones, 14 out of the 15 channels (channels~$21$-$34$) are
\textit{sparsely} used for transmissions.  As shown in Fig.~\ref{loc},
channels allocated to the transmitters are reused inefficiently or at very
large distances.
For instance, out of the $254$ transmitters, only $24$ transmitters in the
four zones operate on channel~$21$.  
% Moreover, only $3$ out of these 24 transmitters are high power
% transmitters. The remaining 21 transmitters are low power transmitters
% with transmission power less than or equal to 500 W.  The coverage area
% of these transmitters are extremely small (less than 5 km), and hence the
% channels could have been reused spatially in a much more efficient way.
We propose a channel allocation scheme such that the minimum number of
TV channels are used in each zone, while ensuring that the coverage areas of
different transmitters do not overlap.  The algorithm of the
proposed channel allocation scheme is as follows.
\begin{algorithm}[h]
% \SetAlgoLined
 %\KwData{List of all TV transmitters, and the transmission powers}
 %\KwResult{Optimal Channel Allocation Scheme}
 \For{All transmitters in the four zones\;}
 {Check if coverage areas of adjacent transmitters overlap\;
 \eIf{Overlap}
 {Check channel numbers of overlapping transmitters\;
 \eIf{Channel numbers are same}
 {Change operating channel of one tower\;
  Calculate coverage area of towers with new operating channels\;
 }{exit}}{exit}}
\end{algorithm}

\begin{figure}[h]
\centering
\includegraphics[trim = 18mm 0mm 18mm 18mm, clip, width=2.5in]{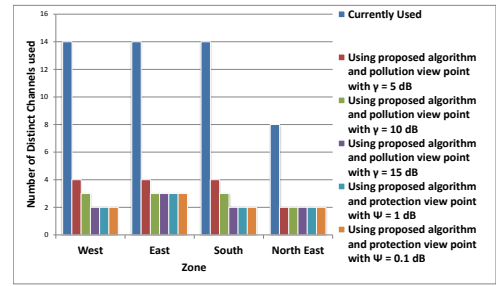}
\caption{Number of channels required in currently and after using the
proposed algorithm}
\label{channel}
\end{figure}

Using the algorithm described above, the minimum number of distinct
channels required without any overlap of the coverage areas for four zones
are given in Fig.~\ref{channel}.  Under this channel allocation scheme, 
the maximum number of distinct TV channels
required in the entire zone is four, which is much smaller than the fourteen
channels currently used in India. To avoid adjacent channel interference, the
overlapping channels must be non-adjacent. 
% With this constraint, the amount of contiguous channels available for
% secondary services would be very high.
%

\section{Conclusions}
\label{sec:conclusions}

In this paper, quantitative analysis of the available TV white space
in the $470$-$590$MHz UHF TV band in India was performed. It is
observed that unlike developed countries, a major portion of TV band
spectrum is unutilized in India.  The results show that even while
using conservative parameters, in at least $36.43$\% areas in the four
zones all the $15$ channels ($100$\% of the TV band spectrum) are
free!  The average available TV white space was calculated using two
methods: (i) the protection and pollution viewpoints~\cite{quant_3};
and, (ii) the FCC regulations~\cite{FCC_2008}. By both methods, the
average available TV white space in the UHF TV band was shown to be
more than $100$MHz!  An algorithm was proposed for reassignment of TV
transmitter frequencies to maximize unused spectrum. It was observed
that four TV channels (or $32$MHz) are sufficient to provide the
existing UHF TV band coverage in India.

In the future, we plan to obtain and include the missing north zone
data in our work. We also wish to explore suitable regulations in
India for the TV white space to enable affordable broadband coverage.
This is timely and important since policy intent for TV white space
was made in NFAP~2011.

\end{document}